\documentclass[12pt]{article}

\usepackage{graphicx}

\newcommand{\ket}[1]{|\, #1 \rangle}
\newcommand{\udt}[3]{#1^{#2}_{\phantom{#2}#3}}
\newcommand{\dut}[3]{#1_{#2}^{\phantom{#2}#3}}
\newcommand{\dudt}[4]{#1_{#2\phantom{#3}#4}^{\phantom{#2}#3}}
\begin{document}
\begin{center}
{\large{\bf Einstein-Podolsky-Rosen correlation \\
in gravitational field}}
\vskip .5 cm
{\bf Hiroaki Terashima$^{1}$ and Masahito Ueda$^{1,2}$}
\vskip .4 cm
{\it $^1$Department of Physics, Tokyo Institute of Technology,\\
Tokyo 152-8551, Japan} \\
{\it $^2$CREST, Japan Science and Technology Corporation (JST),\\
Saitama 332-0012, Japan}
\vskip 0.5 cm
\end{center}

\begin{abstract}
For quantum communication in a gravitational field,
the properties of the
Einstein-Podolsky-Rosen (EPR) correlation
are studied within the framework of general relativity.
Acceleration and gravity
are shown to deteriorate the perfect anti-correlation
of an EPR pair of spins in the \emph{same} direction,
and apparently decrease
the degree of the violation of Bell's inequality.
To maintain the perfect EPR correlation and
the maximal violation of Bell's inequality,
observers must measure the spins
in appropriately chosen \emph{different} directions.
Which directions are appropriate
depends on the velocity of the particles,
the curvature of the spacetime,
and the positions of the observers.
Near the event horizon of a black hole,
the appropriate directions depend
so sensitively on the positions of the observers
that even a very small uncertainty
in the identification of the observers' positions
leads to a fatal error in quantum communication,
unless the observers fall into the black hole
together with the particles.
\end{abstract}

\begin{flushleft}
{\scriptsize
{\bf PACS} : 03.65.Ud, 04.20.-q, 03.67.-a, 02.40.-k \\
{\bf Keywords} : quantum communication, general relativity,
EPR correlation, Bell's inequality
}
\end{flushleft}

\section{Introduction}
Entanglement is a strange feature of quantum theory
and gives rise to a non-local correlation
called the Einstein-Podolsky-Rosen (EPR)
correlation~\cite{EiPoRo35,Bohm89}.
The EPR correlation was originally
discussed to address the foundation of quantum theory.
However, it is now widely recognized as a vital resource
in quantum communication
such as quantum teleportation~\cite{BBCJPW93}
and quantum cryptography~\cite{Ekert91,BeBrMe92}.
Thus, to perform precise quantum communication,
we must understand the properties of the EPR correlation
in various physical situations.
Recently, a number of
articles~\cite{PeScTe02,AlsMil02,TerUed02,RemSmo02,%
GinAda02,ALMH02,LiDu02}
have discussed how entanglement is affected
by the Lorentz transformation
in the regime of special relativity~\cite{Wigner39}.
For example,
the present authors previously discussed
the EPR correlation with moving observers
who share a common rest frame~\cite{TerUed02}.
In that study, we showed that
the perfect anti-correlation of an EPR pair
of spins in the \emph{same} direction deteriorates
in the observers' rest frame,
and that the degree of the violation
of Bell's inequality~\cite{Bell64,CHSH69}
decreases apparently as
the velocity of the observers increases.
To utilize the perfect EPR correlation
and the maximal violation of Bell's inequality,
the moving observers must measure the spins
in appropriately chosen \emph{different} directions;
the choice of such directions depends
on the velocity of the observers
and on that of the particles.

In this paper, we extend these considerations
to a regime of general relativity
by introducing a gravitational field.
In general relativity, a gravitational field is
represented by a curved spacetime, which
entails a breakdown of the global rotational symmetry.
A spin is, on the other hand, known
to represent the rotational symmetry of a system.
Thus, the spin in general relativity can be
defined only locally
by invoking the local rotational symmetry
of the local inertial frame.
In the present paper, we show explicitly how to
extract the \emph{non-local} correlation
from the \emph{locally} defined spins.
We also investigate how to extract
the non-local correlation beyond the event horizon,
which arises in strong gravitational fields
and uniquely features the problem of general relativity.

As a consequence of the local definition,
a particle moving in curved spacetime
is accompanied by a precession of its spin
due to the acceleration of the particle
by an external force
and the difference between
local inertial frames at different points.
These effects of the particle's motion
result in a continuous succession of
local Lorentz transformations.
Since a Lorentz transformation rotates
the spin of the particle
according to the Wigner rotation~\cite{Wigner39},
the motion of the particle leads to
a continuous succession of local Wigner rotations
of the spin, producing spin precession.
Spin precession caused by
acceleration may be viewed as
a generalized Thomas precession in curved spacetime.
On the other hand,
spin precession caused by
a change in the local inertial frame
is an effect of spacetime curvature.

Applying both forms of spin precession
to a relativistic EPR state
near the Schwarzschild black hole,
we show that acceleration and gravity
deteriorate the perfect anti-correlation
in directions that would be the same as each other
if the spacetime were flat,
and that they apparently decrease
the degree of the violation of Bell's inequality,
as in the case of moving observers
in special relativity~\cite{TerUed02}.
To exploit the perfect EPR correlation
and the maximal violation of Bell's inequality,
the observers must measure the spins
in appropriately chosen different directions.
Identification of the appropriate directions
depends on the velocity of the particles,
the curvature of the spacetime, and
the positions of the observers.
Surprisingly,
while the parallel transport in general relativity
can define which directions are the same as each other
in curved spacetime~\cite{vonMen00},
the appropriate directions are \emph{not}
the same as each other even in this sense.
We also show that, near the event horizon of the black hole,
the appropriate directions depend
quite sensitively on the positions of the observers,
because the spin precession
induced by the particle's motion becomes very rapid there.
Therefore, even a very small uncertainty
in the identification of the observers' positions
leads to a fatal error
in quantum communication near the event horizon.
In particular,
static observers cannot extract the EPR correlation
from circularly moving particles
unless they can find their own positions with infinite accuracy.
To exploit the EPR correlation on (and beyond) the event horizon,
the observers must fall into the black hole
together with the particles.

An interesting distinction between the present
general-relativistic problem and the special-relativistic
one~\cite{PeScTe02,AlsMil02,TerUed02,GinAda02,LiDu02}
is that in the former the Lorentz transformation arises from
the motion of the particle, not from that of the observer,
because neither the general coordinate system
nor the local inertial frame is changed at each point.
This means that we could extend previous work on
special-relativistic quantum information
to the regime of general relativity
only by considering moving particles in a gravitational field.
Recently, quantum communications with
accelerated observers have been
discussed~\cite{vanRud03,AlsMil03}
using the Davies-Unruh effect~\cite{Takagi86}
in Minkowski spacetime.
However, their protocols have no special-relativistic analogy,
since they do not involve the Lorentz transformation.
Actually, their resources are not
the entanglement between spins.

This paper is organized as follows.
Section~\ref{sec:spin} formulates
a spin-$1/2$ particle in a curved spacetime
and shows a spin precession induced by 
the motion of the particle.
As an example, the Schwarzschild spacetime is considered.
Section~\ref{sec:eprbell} discusses the EPR correlation and
Bell's inequality in the Schwarzschild spacetime, and
explains how to extract the EPR correlation on and beyond
the event horizon subject to an uncertainty
in the identification of observers' positions.
Section~\ref{sec:summary} summarizes our results.

\section{\label{sec:spin}Spin in Curved Spacetime}

\subsection{Local Inertial Frame}
In general relativity, a gravitational field is represented
by a curved spacetime with metric $g_{\mu\nu}(x)$.
To define a spin in the curved spacetime,
we introduce a local inertial frame at each point
using a vierbein (or a tetrad)
$\dut{e}{a}{\mu}(x)$ and
its inverse $\udt{e}{a}{\mu}(x)$ defined by~\cite{Nakaha90}
\begin{equation}
 \dut{e}{a}{\mu}(x) \, \dut{e}{b}{\nu}(x)
  \,g_{\mu\nu}(x)=\eta_{ab},
\label{eq:defe}
\end{equation}
and
\begin{equation}
 \udt{e}{a}{\mu}(x)\,\dut{e}{a}{\nu}(x)=\dut{\delta}{\mu}{\nu},\qquad
 \udt{e}{a}{\mu}(x)\,\dut{e}{b}{\mu}(x)=\udt{\delta}{a}{b},
\label{eq:defei}
\end{equation}
where $\eta_{ab}=\mathrm{diag}(-1,1,1,1)$ is the Minkowski metric.
Here and henceforth, it is assumed that
Latin letters run over
the four inertial-coordinate labels $0,1,2,3$,
that Greek letters run over
the four general-coordinate labels,
and that repeated indices are to be summed.
The general-coordinate labels are lowered by $g_{\mu\nu}(x)$
and raised by its inverse, $g^{\mu\nu}(x)$,
defined by $g^{\mu\rho}(x)g_{\rho\nu}(x)=\udt{\delta}{\mu}{\nu}$.
The inertial-coordinate labels are lowered by $\eta_{ab}$
and raised by its inverse, $\eta^{ab}$,
defined by $\eta^{ac}\eta_{cb}=\udt{\delta}{a}{b}$.
The vierbein represents
the coordinate transformation
from the general coordinate system $x^\mu$
to the local inertial frame $x^a$ at each point.
Therefore, the vierbein and its inverse transform
a tensor in the general coordinate system
into one in the local inertial frame, and vice versa.
For example, a tensor $\udt{V}{\mu\nu}{\rho}(x)$
in the general coordinate system
can be transformed into that
in the local inertial frame at $x^\mu$ via the relation
$\udt{V}{ab}{c}(x)=\udt{e}{a}{\mu}(x)\udt{e}{b}{\nu}(x)
\dut{e}{c}{\rho}(x)\,\udt{V}{\mu\nu}{\rho}(x)$.
Clearly, the choice of the local inertial frame is not unique,
since the inertial frame remains inertial
under the Lorentz transformation.
The choice of the vierbein therefore has the same degree of
freedom known as the local Lorentz transformation.
In fact, definitions (\ref{eq:defe}) and (\ref{eq:defei})
remain unaltered under the local Lorentz transformation,
i.e.,
\begin{eqnarray}
  \dut{e}{a}{\mu}(x) &\to& \dut{{e'}}{a}{\mu}(x)
    =\dut{\Lambda}{a}{b}(x) \,\dut{e}{b}{\mu}(x), \\
  \udt{e}{a}{\mu}(x) &\to& \udt{{e'}}{a}{\mu}(x)
    =\udt{\Lambda}{a}{b}(x) \,\udt{e}{b}{\mu}(x),
\end{eqnarray}
where
$\dut{\Lambda}{a}{b}(x)=\eta_{ac}\,\eta^{bd}\,\udt{\Lambda}{c}{d}(x)$
and
$\udt{\Lambda}{a}{c}(x)\udt{\Lambda}{b}{d}(x)\,\eta^{cd}=\eta^{ab}$.
Although this transformation $\udt{\Lambda}{a}{b}(x)$
is a Lorentz transformation,
it has no connection with the Lorentz transformation
that is included as a special case
in the general coordinate transformation.

Using the local Lorentz transformation,
we can define a particle with spin $1/2$
in curved spacetime.
It is well known that a ``particle''
is not defined uniquely in quantum field theory
in curved spacetime~\cite{BirDav82},
since the ``time'' coordinate
to define the positive energy is not unique;
this non-uniqueness is an origin of particle creation
such as Hawking radiation~\cite{Hawkin75}.
However, in the present formulation,
our particle is specified
by the vierbein $\dut{e}{0}{\mu}(x)$, which
relates the local time to a global time.
A spin-$1/2$ particle in curved spacetime
is then defined as a particle whose one-particle states
furnish the spin-$1/2$ representation
of the local Lorentz transformation,
not of the general coordinate transformation.
In fact, the Dirac field
in the curved spacetime~\cite{BirDav82}
is spinor under the local Lorentz transformation,
while it is scalar
under the general coordinate transformation.
More specifically,
consider a massive spin-$1/2$ particle moving with
four-velocity $u^\mu(x)=dx^\mu/d\tau$, which is normalized as
\begin{equation}
u^\mu(x)\,u_\mu(x)=-c^2.
\label{eq:defu}
\end{equation}
The four-momentum is given by $p^\mu(x)=mu^\mu(x)$,
with $m$ being the mass of the particle.
Accordingly,
the four-momentum in the local inertial frame becomes
$p^a(x)=\udt{e}{a}{\mu}(x)\, p^\mu(x)$ using the vierbein.
In the local inertial frame at point $x^\mu$,
the one-particle state in quantum theory is specified
by the third-component $\sigma$
(=$\uparrow$, $\downarrow$) of the spin
as $\ket{p^a(x),\sigma\,;x}$, as in special relativity.
This state indicates not
a localized state at $x^\mu$ with definite momentum $p^a(x)$,
but rather an extended state whose momentum is $p^a(x)$
if it is viewed in the local inertial frame at $x^\mu$.
By definition,
the state $\ket{p^a(x),\sigma\,;x}$ transforms
as the spin-$1/2$ representation
under the local Lorentz transformation.
Note that, in the case of special relativity,
a one-particle state $\ket{p^a,\sigma}$
in the spin-$1/2$ representation transforms
under a Lorentz transformation $\udt{\Lambda}{a}{b}$
as~\cite{Weinbe95,Ohnuki88}
\begin{equation}
 U(\Lambda)\, \ket{p^a,\sigma}
  = \sum_{\sigma'} D_{\sigma'\sigma}^{(1/2)}(W(\Lambda,p))
    \,\ket{\Lambda p^a,\sigma'},
\end{equation}
where $D_{\sigma'\sigma}^{(1/2)}(W(\Lambda,p))$ is
a $2\times 2$ unitary matrix that rotates
the spin of the particle according to
the Wigner rotation $\udt{W}{a}{b}(\Lambda,p)$.
The explicit form of the Wigner rotation is
given by
\begin{equation}
\udt{W}{a}{b}(\Lambda,p)=
\left[L^{-1}(\Lambda p)\,\Lambda\,L(p)\right]\udt{}{a}{b}
\label{eq:defw}
\end{equation}
with a standard Lorentz transformation $\udt{L}{a}{b}(p)$,
\begin{eqnarray}
\udt{L}{0}{0}(p) &=& \gamma, \nonumber \\
\udt{L}{0}{i}(p) &=& \udt{L}{i}{0}(p)=p^i/mc,    \\
\udt{L}{i}{k}(p) &=& \delta_{ik}+
(\gamma-1)\,p^i\,p^k/|\vec{p}|^2,\nonumber
\end{eqnarray}
where $\gamma=\sqrt{|\vec{p}|^2+m^2c^2}/mc$ and $i,k=1,2,3$.
Therefore, in the case of the curved spacetime,
the one-particle state $\ket{p^a(x),\sigma\,;x}$ transforms
under a local Lorentz transformation $\udt{\Lambda}{a}{b}(x)$ as
\begin{equation}
 U(\Lambda(x))\, \ket{p^a(x),\sigma;x}
  = \sum_{\sigma'} D_{\sigma'\sigma}^{(1/2)}(W(x))
    \,\ket{\Lambda p^a(x),\sigma';x},
    \label{eq:ll}
\end{equation}
where $\udt{W}{a}{b}(x)\equiv\udt{W}{a}{b}(\Lambda(x),p(x))$
is the local Wigner rotation.

Instead of the one-particle state $\ket{p^a(x),\sigma\,;x}$,
we could employ Dirac's four-component spinor,
which also represents a spin-$1/2$ particle.
These two are equivalent as far as
Lorentz-transformation properties are concerned.
For example,
the case of moving observers~\cite{TerUed02}
in special relativity was calculated
equivalently using Dirac's spinor~\cite{AlsMil02}.

\subsection{Spin Precession}
Since the spin of a particle is defined locally
relative to the local inertial frame,
we next consider the change of the spin when
a particle moves from one point to another in curved spacetime.
After an infinitesimal proper time $d\tau$,
the particle moves to a new point
$x'^\mu=x^\mu+u^\mu(x)\,d\tau$.
The four-momentum of the particle then becomes
$p^a(x')=p^a(x)+\delta p^a(x)$ in the local inertial frame
at the new point,
because of changes in both momentum and local inertial frame:
\begin{equation}
 \delta p^a(x) = \delta p^\mu(x)\, \udt{e}{a}{\mu}(x)
  + p^\mu(x) \,\delta\udt{e}{a}{\mu}(x).
\label{eq:delpa}
\end{equation}
The change in the momentum is given by
\begin{equation}
  \delta p^\mu(x) =
    u^\nu(x)\, d\tau \nabla_\nu p^\mu(x)
   = m a^\mu(x) \,d\tau,
\end{equation}
where
\begin{equation}
 a^\mu (x)=u^\nu(x)\nabla_\nu u^\mu(x)
\label{eq:defa}
\end{equation}
is the acceleration
due to an external force (excluding the gravity).
Since
$p^\mu(x)\, p_\mu(x)=-m^2c^2$ and $p^\mu(x)\, a_\mu(x)=0$
from Eq.~(\ref{eq:defu}), we obtain
\begin{equation}
 \delta p^\mu(x)=
    -\frac{1}{mc^2}\left[\,a^\mu(x)\,p_\nu(x)-
       p^\mu(x)\,a_\nu(x)\,\right]\,p^\nu(x)\,d\tau.
\label{eq:delpmu}
\end{equation}
On the other hand,
the change in the local inertial frame is given by
\begin{eqnarray}
  \delta\udt{e}{a}{\mu}(x) &=&
    u^\nu(x)\, d\tau \nabla_\nu\udt{e}{a}{\mu}(x) \nonumber \\
   &=& -u^\nu(x)\,\dudt{\omega}{\nu}{a}{b}(x)
       \, \udt{e}{b}{\mu}(x)\, d\tau  \nonumber \\
  &\equiv & \udt{\chi}{a}{b}(x)\, \udt{e}{b}{\mu}(x)\, d\tau,
\label{eq:defchi}
\end{eqnarray}
where
\begin{equation}
\dudt{\omega}{\mu}{a}{b}(x)
=-\dut{e}{b}{\nu}(x)\nabla_\mu \udt{e}{a}{\nu}(x)
=\udt{e}{a}{\nu}(x)\nabla_\mu \dut{e}{b}{\nu}(x)
\label{eq:defome}
\end{equation}
is the connection one-form
(or a spin connection)~\cite{Nakaha90}.
The second equality in Eq.~(\ref{eq:defome}) results
from definition (\ref{eq:defe}) and
$\nabla_\mu g_{\nu\rho}(x) = \nabla_\mu \eta_{ab}=0$,
giving $\chi_{ab}(x)=-\chi_{ba}(x)$.
Substituting Eqs.~(\ref{eq:delpmu}) and (\ref{eq:defchi})
into Eq.~(\ref{eq:delpa}), we obtain
\begin{equation}
 \delta p^a(x) = \udt{\lambda}{a}{b}(x) \,p^b(x)\, d\tau,
\end{equation}
where
\begin{equation}
  \udt{\lambda}{a}{b}(x)=
-\frac{1}{mc^2}\left[\,a^a(x)\,p_b(x)-p^a(x)\,a_b(x)\,\right]
+\udt{\chi}{a}{b}(x).
\label{eq:deflam}
\end{equation}
This is an infinitesimal local Lorentz transformation,
since $\lambda_{ab}(x)=-\lambda_{ba}(x)$.
That is, when the particle moves,
the momentum in the local inertial frame transforms
under the local Lorentz transformation
\begin{equation}
  \udt{\Lambda}{a}{b}(x)=\udt{\delta}{a}{b}
  +\udt{\lambda}{a}{b}(x)\, d\tau.
\label{eq:ill}
\end{equation}

Using the unitary operator corresponding to
this local Lorentz transformation,
the state $\ket{p^a(x),\sigma\,;x}$ is now described as
$U(\Lambda(x))\,\ket{p^a(x),\sigma\,;x'}$
in the local inertial frame at the new point $x'^\mu$.
From Eq.~(\ref{eq:ll}),
the spin is then rotated according to
the local Wigner rotation $\udt{W}{a}{b}(x)$.
For the infinitesimal Lorentz transformation (\ref{eq:ill}),
the infinitesimal Wigner rotation becomes
\begin{equation}
 \udt{W}{a}{b}(x)=\udt{\delta}{a}{b}
   +\udt{\vartheta}{a}{b}(x)\, d\tau,
   \label{eq:iwr}
\end{equation}
where
$\udt{\vartheta}{0}{0}(x)=\udt{\vartheta}{0}{i}(x)
=\udt{\vartheta}{i}{0}(x)=0$ and
\begin{equation}
 \udt{\vartheta}{i}{k}(x)=\udt{\lambda}{i}{k}(x)
   +\frac{\udt{\lambda}{i}{0}(x)\,p_k(x)-
   \lambda_{k0}(x)\,p^i(x)}{p^0(x)+mc}.
 \label{eq:deflw}
\end{equation}
Its spin-$1/2$ representation is
\begin{eqnarray}
 D_{\sigma'\sigma}^{(1/2)}(W(x)) &=&
  I+\frac{i}{2}\bigl[\,\vartheta_{23}(x)\, \sigma_x
   +\vartheta_{31}(x)\, \sigma_y
  \nonumber \\
 & &\qquad{}+
  \vartheta_{12}(x)\, \sigma_z \,\bigr]\,d\tau,
\end{eqnarray}
with the Pauli matrices $\{\sigma_x, \sigma_y, \sigma_z\}$
and the unit matrix $I$.
We thus obtain the formula
for $\ket{p^a(x),\sigma\,;x}=U(\Lambda(x))\,\ket{p^a(x),\sigma\,;x'}$ as
\begin{eqnarray}
&& U(\Lambda(x))\,\ket{p^a(x),\uparrow\,;x'} \nonumber \\
&& \qquad =\left(1+\frac{i}{2}\vartheta_{12}(x)\,d\tau\right)
     \,\ket{p^a(x'),\uparrow\,;x'} \nonumber \\
&& \qquad\quad{}-\frac{1}{2}
     \left(\vartheta_{31}(x)-i\vartheta_{23}(x)\right)\,d\tau
     \, \ket{p^a(x'),\downarrow\,;x'}, \\
&& U(\Lambda(x))\,\ket{p^a(x),\downarrow\,;x'} \nonumber \\
&& \qquad=\frac{1}{2}\left(\vartheta_{31}(x)+i\vartheta_{23}(x)\right)\,d\tau
    \, \ket{p^a(x'),\uparrow\,;x'} \nonumber \\
&&  \qquad\quad{}+\left(1-\frac{i}{2}\vartheta_{12}(x)\,d\tau\right)
     \, \ket{p^a(x'),\downarrow\,;x'}.
\end{eqnarray}
We emphasize that the change of the spin $\udt{\vartheta}{a}{b}(x)$
is equal neither to $\udt{\chi}{a}{b}(x)$
nor to $\udt{\lambda}{a}{b}(x)$.

By iterating the infinitesimal transformation,
we find a transformation formula for a finite proper time.
This becomes a Dyson series as
in the time-dependent perturbation theory,
since $\udt{\lambda}{a}{b}(x)$'s at different points
do not necessarily commute with each other.
Suppose that the particle moves along
a path $x^\mu(\tau)$ from $x_i^\mu=x^\mu(\tau_i)$
to $x_f^\mu=x^\mu(\tau_f)$.
Dividing $h=\tau_f-\tau_i$ into $N$ parts with
$x^\mu_{(k)}\equiv x^\mu(\tau_i+(hk/N))$
and applying the infinitesimal
Lorentz transformation (\ref{eq:ill}) to each part,
we obtain the Lorentz transformation for the momentum
in the local inertial frame as
\begin{eqnarray}
 \udt{\Lambda}{a}{b}(x_f,x_i) &=&
  \lim_{N\to\infty}
  \prod_{k=0}^{N} \left[ \udt{\delta}{a}{b}
  +\udt{\lambda}{a}{b}(x_{(k)})\, \frac{h}{N}\right] \nonumber \\
   &=&T
  \exp\left[\int^{\tau_f}_{\tau_i}
  \udt{\lambda}{a}{b}(x(\tau))\, d\tau\right],
\end{eqnarray}
where $T$ is the time-ordering operator, and
the exponential refers not to the exponential of each component
but to that of the whole matrix.
The corresponding Wigner rotation is then given by
\begin{eqnarray}
  \udt{W}{a}{b}(x_f,x_i)&=&
  \lim_{N\to\infty}
  \prod_{k=0}^{N} \left[ \udt{\delta}{a}{b}
  +\udt{\vartheta}{a}{b}(x_{(k)})\, \frac{h}{N}\right] \nonumber \\
    &=&T
   \exp\left[\int^{\tau_f}_{\tau_i}
    \udt{\vartheta}{a}{b}(x(\tau))\, d\tau\right].
  \label{eq:fwr}
\end{eqnarray}
This formula can be proven by noting that
\begin{equation}
\udt{W}{a}{b}(\Lambda_1\Lambda_2,p)
=\left[\,W(\Lambda_1,\Lambda_2p)
\,W(\Lambda_2,p)\,\right]\udt{}{a}{b}
\end{equation}
from the definition (\ref{eq:defw}), and that
\begin{equation}
\left[ \udt{\delta}{a}{b}
+\udt{\lambda}{a}{b}(x_{(k)})\,\frac{h}{N}\right]\,
p^b(x_{(k)})=p^a(x_{(k+1)})
\end{equation}
from the definition of $\udt{\lambda}{a}{b}(x)$.

\subsection{Schwarzschild Spacetime}
As a unique example in general relativity,
we consider the Schwarzschild spacetime~\cite{Wald84},
which is the unique spherically symmetric solution
of Einstein's equation in vacuum.
In the spherical coordinate system $(t,r,\theta,\phi)$,
the metric of this spacetime is given by
\begin{eqnarray}
ds^2=g_{\mu\nu}(x)dx^\mu dx^\nu  &=&
-f(r)\,c^2dt^2+\frac{1}{f(r)}\,dr^2 \nonumber \\
 & &{}+
r^2(d\theta^2+\sin^2\theta d\phi^2)
\end{eqnarray}
where
\begin{equation}
 f(r)=1-\frac{r_s}{r},
\end{equation}
and the parameter $r_s$ is called the Schwarzschild radius.
At this radius $r=r_s$,
the Schwarzschild spacetime has an event horizon
where no displacement $dx^\mu=(dt,dr,d\theta,d\phi)$
can be timelike $ds^2<0$ because of $f(r_s)=0$.
The singularity of the metric at the event horizon
means not a physical singularity but
a breakdown of the coordinate system $(t,r,\theta,\phi)$.
The inside of the event horizon $r<r_s$
is the Schwarzschild black hole
whose mass is given by $M=c^2r_s/2G$.
In the Schwarzschild spacetime,
it is convenient to choose vierbein (\ref{eq:defe})
such that
\begin{eqnarray}
\dut{e}{0}{t}(x)=\frac{1}{c\sqrt{f(r)}}, &\quad&
\dut{e}{1}{r}(x)=\sqrt{f(r)}, \nonumber \\
\dut{e}{2}{\theta}(x)=\frac{1}{r}, &\quad&
\dut{e}{3}{\phi}(x)=\frac{1}{r\sin\theta},
\label{eq:schvb}
\end{eqnarray}
and all the other components are zero.
In the following discussions,
only non-zero components will be shown.
At each point,
the $0$-, $1$-, $2$-, and $3$-axes are parallel to
the $t$-, $r$-, $\theta$-, and $\phi$-directions,
respectively.
This vierbein represents
a static local inertial frame at each point,
because all the components are independent of $t$ and
because the components $\dut{e}{i}{t}(x)$ ($i=1,2,3$) and
$\dut{e}{0}{\alpha}(x)$ ($\alpha=r,\theta,\phi$) are zero.
The inverse of vierbein (\ref{eq:defei}) is then given by
\begin{eqnarray}
\udt{e}{0}{t}(x)=c\sqrt{f(r)}, &\quad&
\udt{e}{1}{r}(x)=\frac{1}{\sqrt{f(r)}},\nonumber \\
\udt{e}{2}{\theta}(x)=r, &\quad&
\udt{e}{3}{\phi}(x)=r\sin\theta.
\end{eqnarray}
A straightforward calculation shows that
the connection one-form (\ref{eq:defome}) becomes
\begin{eqnarray}
\dudt{\omega}{t}{0}{1}(x) &=&
     \dudt{\omega}{t}{1}{0}(x)=\frac{cr_s}{2r^2}, \\
\dudt{\omega}{\theta}{1}{2}(x) &=&
    -\dudt{\omega}{\theta}{2}{1}(x)=-\sqrt{f(r)}, \\
\dudt{\omega}{\phi}{1}{3}(x) &=&
    -\dudt{\omega}{\phi}{3}{1}(x)=-\sqrt{f(r)}\,\sin\theta, \\
\dudt{\omega}{\phi}{2}{3}(x) &=&
    -\dudt{\omega}{\phi}{3}{2}(x)=-\cos\theta.
\end{eqnarray}

In this Schwarzschild spacetime,
let us consider a particle in a circular motion
with a radius $r$ ($>r_s$) and constant velocity
$rd\phi/dt\equiv v\sqrt{f(r)}$
on the equatorial plane $\theta=\pi/2$.
The four-velocity of this particle is given by
\begin{equation}
  u^{t}(x)=\frac{\cosh\xi}{\sqrt{f(r)}},\qquad
  u^{\phi}(x)=\frac{c\,\sinh\xi}{r},
\label{eq:schfv}
\end{equation}
where $\xi$ is a rapidity in the local inertial frame
defined by
\begin{equation}
 \frac{v}{c}=\tanh\xi.
\end{equation}
In order for the particle to move in this way,
we must apply an external force against the centrifugal force
and the gravity.
The acceleration (\ref{eq:defa})
due to this external force then becomes
\begin{equation}
 a^r(x)=-\frac{c^2\sinh^2\xi}{r}
   \left[1-\frac{r_s}{2rf(r)}\coth^2\xi\right]f(r).
\end{equation}

\begin{figure}
\begin{center}
\includegraphics[scale=0.6]{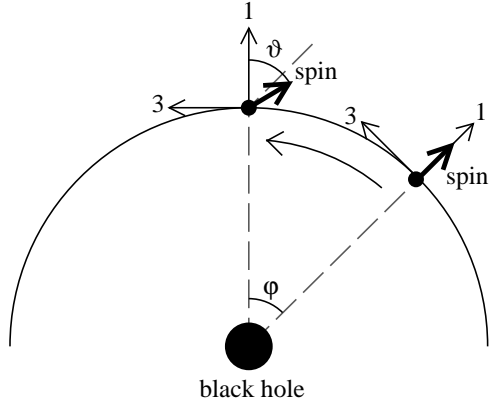}
\end{center}
\caption{\label{fig1}A spin in a circular motion.
At each point,
the $1$- and $3$-axes are parallel to
the $r$- and $\phi$-directions, respectively.}
\end{figure}

Now, after an infinitesimal proper time $d\tau$,
the particle moves by an angle
$\delta\phi=u^\phi(x)d\tau$
as depicted in Fig.~\ref{fig1}.
However,
the change in the local inertial frame
(\ref{eq:defchi}) is \emph{not} a trivial rotation
about the $2$-axis,
\begin{equation}
\udt{\varphi}{1}{3}(x) =
-\udt{\varphi}{3}{1}(x) = u^\phi(x)=
\frac{c\sinh\xi}{r},
\label{eq:trirot}
\end{equation}
since in general relativity a parallel transport
that depends on the spacetime curvature
is required to compare local inertial frames at different points.
The definition (\ref{eq:defchi}) shows that
the change in the local inertial frame
consists of a boost along the $1$-axis and
a rotation about the $2$-axis,
\begin{eqnarray}
\udt{\chi}{0}{1}(x) &=&
\udt{\chi}{1}{0}(x) =
-\frac{cr_s\cosh\xi}{2r^2\sqrt{f(r)}}, \\
\udt{\chi}{1}{3}(x) &=&
-\udt{\chi}{3}{1}(x) =
\frac{c\sinh\xi\sqrt{f(r)}}{r}.
\end{eqnarray}
The infinitesimal Lorentz transformation (\ref{eq:deflam})
then becomes
\begin{eqnarray}
\udt{\lambda}{0}{1}(x) &=& \udt{\lambda}{1}{0}(x)
  \nonumber \\
  &=&-\frac{c\cosh\xi\sinh^2\xi}{r}
   \left[1-\frac{r_s}{2rf(r)}\right]\sqrt{f(r)}, \\
\udt{\lambda}{1}{3}(x) &=& -\udt{\lambda}{3}{1}(x)
 \nonumber \\
  &=&\frac{c\cosh^2\xi\sinh\xi}{r}
   \left[1-\frac{r_s}{2rf(r)}\right]\sqrt{f(r)},
\end{eqnarray}
which also consists of a boost along the $1$-axis and
a rotation about the $2$-axis.
Nevertheless, the momentum in the local inertial frame is
constant, $p^a(x)=(mc\cosh\xi,0,0,mc\sinh\xi)$,
pointing to the $3$-axis.
Finally,
the change of the spin defined by Eq.~(\ref{eq:deflw}) becomes
the rotation about the $2$-axis through an angle
\begin{equation}
\udt{\vartheta}{1}{3}(x) = -\udt{\vartheta}{3}{1}(x)
=\frac{c\cosh\xi\sinh\xi}{r}
   \left[1-\frac{r_s}{2rf(r)}\right]\sqrt{f(r)},
 \label{eq:schwig}
\end{equation}
as illustrated in Fig.~\ref{fig1}.
It is important to note that
\begin{equation}
\udt{\vartheta}{a}{b}(x)\neq\udt{\lambda}{a}{b}(x)
\neq\udt{\chi}{a}{b}(x)\neq\udt{\varphi}{a}{b}(x);
\label{eq:noneq}
\end{equation}
these three non-equalities result from
the boost part of $\udt{\lambda}{a}{b}(x)$,
the acceleration of the particle,
and the curvature of the spacetime, respectively.

To illustrate this result,
we consider a special case, $r_s=0$,
i.e. Minkowski spacetime.
In this case,
the change in the local inertial frame is reduced to
the rotation about the $2$-axis through an angle
\begin{equation}
\udt{\chi}{1}{3}(x)=-\udt{\chi}{3}{1}(x)=\frac{c\sinh\xi}{r},
\label{eq:minchi}
\end{equation}
which is nothing but the trivial rotation (\ref{eq:trirot}).
The change of the spin is also reduced to the rotation
about the $2$-axis through an angle
\begin{equation}
\udt{\vartheta}{1}{3}(x)=-\udt{\vartheta}{3}{1}(x)
=\frac{c\cosh\xi\sinh\xi}{r}.
\label{eq:minwig}
\end{equation}
The difference between (\ref{eq:minchi}) and (\ref{eq:minwig})
gives rise to a precession of the spin
called the Thomas precession.
When $v/c\ll 1$,
the precession angle per $dt=d\tau\cosh\xi$ becomes
\begin{equation}
\left[ \,\udt{\vartheta}{3}{1}(x)-\udt{\chi}{3}{1}(x)\, \right] d\tau
\sim -\frac{va}{2c^2}\,dt,
\end{equation} 
where $a\equiv |a^r(x)|=c^2\sinh^2\xi/r$.

\section{\label{sec:eprbell}EPR Correlation and Bell's Inequality}
We now discuss how to exploit the EPR correlation for
quantum communication using accelerated particles
in a gravitational field.
Specifically, we consider a pair of
circularly moving particles
in the Schwarzschild spacetime,
as discussed in the preceding section.

\subsection{EPR Correlation}
\begin{figure}
\begin{center}
\includegraphics[scale=0.52]{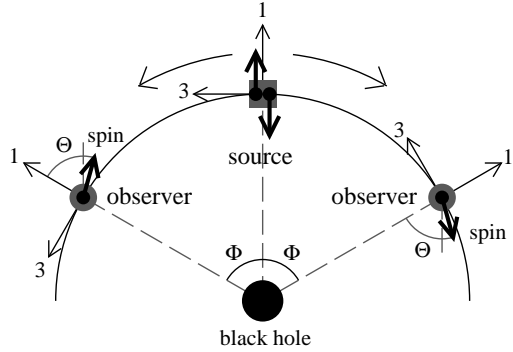}
\end{center}
\caption{\label{fig2}An EPR gedankenexperiment
in the Schwarzschild spacetime.
Two observers (gray circles) and
an EPR source (gray square) are located
at $\phi=\pm\Phi$ and $0$, respectively.}
\end{figure}

Consider two observers and an EPR source
on the equatorial plane $\theta=\pi/2$
at a radius $r$ ($>r_s$) with azimuthal angles
$\pm\Phi$ (observers) and $0$ (EPR source),
as illustrated in Fig.~\ref{fig2}.
The observers and the EPR source are assumed to
be static (``at rest'' in the coordinate system
$(t,r,\theta,\phi)$) and to use the static local inertial frame
(\ref{eq:schvb}) to measure or prepare the spin state.
Note that the inertial frame is defined at each instant,
since the observers and EPR source are accelerated
to stay at a constant radius.
First, the EPR source emits a pair of entangled particles
in opposite directions with constant four-momenta
$p^a_\pm=(mc\cosh\xi,0,0,\pm mc\sinh\xi)$
in the spin-singlet state,
\begin{equation}
\frac{1}{\sqrt{2}}\Bigl[\,
\ket{p^a_+,\uparrow\,;0}\ket{p^a_-,\downarrow\,;0}
-\ket{p^a_+,\downarrow\,;0}\ket{p^a_-,\uparrow\,;0}\,\Bigr],
\end{equation}
where for notational simplicity
we write only the $\phi$-coordinate in the arguments.
After a proper time $r\Phi/c\sinh\xi$,
each particle reaches the corresponding observer.
Using Eq.~(\ref{eq:schwig}),
the Wigner rotation (\ref{eq:fwr}) becomes
a rotation about the $2$-axis
\begin{equation}
  \udt{W}{a}{b}(\pm\Phi,0) =
  \left(
   \begin{array}{cccc}
     1 & 0 & 0 & 0 \\
     0 & \cos\Theta & 0 & \pm\sin\Theta \\
     0 & 0 & 1 & 0 \\
     0 & \mp\sin\Theta & 0 & \cos\Theta
   \end{array}
  \right),
\end{equation}
where the angle $\Theta$ is given by
\begin{equation}
\Theta=\Phi\cosh\xi
 \left[1-\frac{r_s}{2rf(r)}\right]\sqrt{f(r)}.
\end{equation}
Note that we do not need the
time-ordering operator $T$ in this example,
since $\udt{\vartheta}{a}{b}(x)$ is constant
during the motion.
This Wigner rotation is represented
by using the Pauli matrix $\sigma_y$ as
\begin{equation}
 D_{\sigma'\sigma}^{(1/2)}(W(\pm\Phi,0))
   =\exp\left(\mp i\frac{\sigma_y}{2}\Theta\right).
\end{equation}
Therefore,
in the local inertial frames at $\phi=\Phi$ and $-\Phi$,
the state is described as
\begin{eqnarray}
& & \frac{1}{\sqrt{2}}\Biggl[\,\cos\Theta\Bigl( \,
  \ket{p^a_+,\uparrow\,;\Phi}\ket{p^a_-,\downarrow\,;-\Phi}
  \nonumber \\
& & \qquad\qquad\qquad
  {}-\ket{p^a_+,\downarrow\,;\Phi}\ket{p^a_-,\uparrow\,;-\Phi}
  \, \Bigr) \nonumber \\
& & \qquad{}+\sin\Theta\,\Bigl( \,
  \ket{p^a_+,\uparrow\,;\Phi}\ket{p^a_-,\uparrow\,;-\Phi}
 \nonumber \\
& & \qquad\qquad\qquad
  {}+\ket{p^a_+,\downarrow\,;\Phi}\ket{p^a_-,\downarrow\,;-\Phi}
  \, \Bigr) \,\Biggr],
\end{eqnarray}
as illustrated in Fig.~\ref{fig2}.
Because the spin-singlet state is mixed
with the spin-triplet state,
spin measurements in the same direction
are not always anti-correlated
in the local inertial frames at $\phi=\pm\Phi$
(e.g. in each $1$-axis).
Clearly, this result includes
the trivial rotation of the local inertial frames $\pm\Phi$,
as in Eq.~(\ref{eq:trirot}).
To eliminate this spurious effect,
we rotate the bases at $\phi=\pm\Phi$
about the $2$-axis through the angles $\mp\Phi$,
respectively:
\begin{eqnarray}
\ket{p^a_\pm,\uparrow\,;\pm\Phi}' &=&
    \cos\frac{\Phi}{2} \ket{p^a_\pm,\uparrow\,;\pm\Phi}\nonumber \\
 & & \qquad {}\pm\sin\frac{\Phi}{2} \ket{p^a_\pm,\downarrow\,;\pm\Phi}, \\
  \ket{p^a_\pm,\downarrow\,;\pm\Phi}' &=&
   \mp \sin\frac{\Phi}{2} \ket{p^a_\pm,\uparrow\,;\pm\Phi}\nonumber \\
 & & \qquad {}+\cos\frac{\Phi}{2} \ket{p^a_\pm,\downarrow\,;\pm\Phi}.
\end{eqnarray}
Using these bases, the state is written as
\begin{eqnarray}
& & \frac{1}{\sqrt{2}} \Biggl[\,\cos\Delta\Bigl( \, 
  \ket{p^a_+,\uparrow\,;\Phi}'\ket{p^a_-,\downarrow\,;-\Phi}' \nonumber \\
& & \qquad\qquad\qquad{}
  -\ket{p^a_+,\downarrow\,;\Phi}'\ket{p^a_-,\uparrow\,;-\Phi}'
  \, \Bigr) \nonumber \\
 & & \qquad{}+\sin\Delta\,\Bigl( \,
  \ket{p^a_+,\uparrow\,;\Phi}'\ket{p^a_-,\uparrow\,;-\Phi}' \nonumber \\
& & \qquad\qquad\qquad{}
 +\ket{p^a_+,\downarrow\,;\Phi}'\ket{p^a_-,\downarrow\,;-\Phi}'
  \, \Bigr)\,\Biggr],
\end{eqnarray}
where
\begin{equation}
  \Delta=\Theta-\Phi=\Phi\left\{\cosh\xi
 \left[1-\frac{r_s}{2rf(r)}\right]\sqrt{f(r)}-1\right\}.
\end{equation}
Since the trivial rotation is removed in these bases,
we can explicitly see that the relativistic effect
deteriorates the perfect anti-correlation in the directions
that would be the same as each other
if the spacetime were flat.
This deterioration is a consequence of the non-equality
$\udt{\vartheta}{a}{b}(x)\neq\udt{\varphi}{a}{b}(x)$
in Eq.~(\ref{eq:noneq}).
Of course, our result does not mean
a breakdown of the non-local correlation,
since the entanglement is invariant
under local unitary operations.
If we take account of the relativistic effect
arising from acceleration and gravity,
we can exploit the perfect anti-correlation
for quantum communication.
More specifically,
the observers at $\phi=\pm\Phi$ must rotate
the directions of the measurement about the $2$-axis
through the angles $\mp\Theta$ in their local inertial frames,
respectively.
It is interesting that
the parallel transport in general relativity~\cite{vonMen00}
does \emph{not} give the directions that
maintain the perfect anti-correlation.
This occurs because of
$\udt{\vartheta}{a}{b}(x)\neq\udt{\chi}{a}{b}(x)$
in Eq.~(\ref{eq:noneq}).

\begin{figure}
\begin{center}
\includegraphics[scale=0.65]{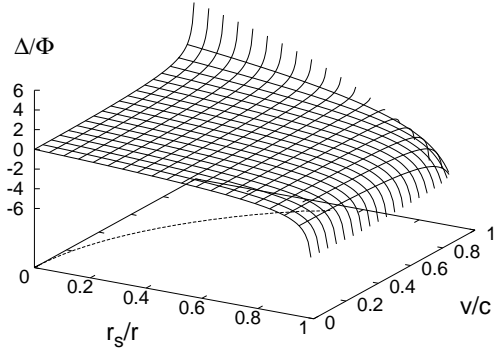}
\end{center}
\caption{\label{fig3}The angle $\Delta$ as a function
of $r_s/r$ and $v/c=\tanh\xi$.
The dotted line at the bottom is the radius $r_0$,
where $\Delta$ becomes $0$.
$|\Delta/\Phi|$ is on the order of $10^{-9}$
on the earth if $v\ll 10$ km/s.
However, it becomes infinite as $r\to r_s$ or $v\to c$.}
\end{figure}

The value of $\Delta$ as a function of $r_s/r$ and $v/c=\tanh\xi$
is shown in Fig.~\ref{fig3}.
In the non-relativistic limit $v/c\to0$ and $r_s/r\to0$,
the angle $\Delta$ becomes
\begin{equation}
  \Delta\sim \Phi\left(\frac{v^2}{2c^2}-\frac{r_s}{r}\right).
\label{eq:nonrela}
\end{equation}
The first term is attributable to acceleration and
the second to gravity.
Note that they have different signs.
Although Eq.~(\ref{eq:nonrela}) holds only
for the non-relativistic limit,
we can draw from it
the following qualitative physical picture:
At the spatial infinity $r\to\infty$,
the gravitational field is so weak that
the angle $\Delta$ is positive.
However, closer to the event horizon,
the gravitational field becomes stronger,
thus making $\Delta$ smaller.
At a radius $r=r_0$ defined by
\begin{equation}
 \left[1-\frac{r_s}{2r_0f(r_0)}\right]\sqrt{f(r_0)}
   =\sqrt{1-\frac{v^2}{c^2}},
\end{equation}
the angle $\Delta$ vanishes,
and becomes negative for $r<r_0$ (see Fig.~\ref{fig3}).
In the limit of $v\to c$,
the radius $r_0$ becomes $3r_s/2$;
inside this radius,
the gravitational field is so strong that
the acceleration $a^r(x)$ for the circular motion
must be in the outward direction for any $v$.
Right on the event horizon $r\to r_s$,
we find $\Delta\to-\infty$.

\subsection{Uncertainties in Observers' Positions}
We have shown that the observers can,
in principle, utilize the EPR correlation
by adjusting the directions of measurement.
Nevertheless, in practice,
such adjustments become difficult near the horizon,
since the spin precession (\ref{eq:schwig})
is very rapid there.
Suppose a classical or quantum uncertainty
$\delta\Phi$ exists in the observers' position $\Phi$.
The error of the angle $\Theta$
to maintain the perfect EPR correlation then becomes
\begin{equation}
 \delta\Theta=\delta\Phi\left|
1+\frac{\Delta}{\Phi}\right|.
\label{eq:err}
\end{equation}
Quantum communication must tolerate this error by
some error-correcting scheme.
However, near the horizon $r\sim r_s$,
$\delta\Theta$ can be much larger than $\pi$
and thus the observers cannot determine
the directions of measurement
clearly enough to extract the EPR correlation.
Therefore, to utilize the EPR correlation,
$\delta\Phi$ and $r$ must satisfy at least
\begin{equation}
 \delta\Phi<\pi
 \left|1+\frac{\Delta}{\Phi}\right|^{-1}.
\end{equation}
At the horizon $r=r_s$, this requirement
reduces to $\delta\Phi=0$
because the velocity of the spin precession
(\ref{eq:schwig}) is infinite.
This means that the observers on the horizon could not
extract the EPR correlation from the particles,
were it not for the infinite accuracy of $\Phi$.
More generally,
for a given uncertainty $\delta\Phi$,
there exists a radius $r_c$ ($>r_s$) such that
static observers at $r<r_c$ cannot extract
the EPR correlation from circularly moving particles.

However, if the observers use different
local inertial frames and different particles,
they can extract the EPR correlation even at $r<r_c$.
Note that the divergence of
the spin precession (\ref{eq:schwig})
comes from the fact that
the vierbein (\ref{eq:schvb})
and four-velocity (\ref{eq:schfv}) become singular
at the horizon.
Therefore, the observers must choose
a vierbein and a four-velocity
that avoid the singularities at the horizon.
Since these singularities are connected
with the breakdown of
the coordinate system $(t,r,\theta,\phi)$,
we adopt the Kruskal coordinate system~\cite{Wald84},
in which the metric is not singular at the horizon.
The Kruskal coordinates $(T,R)$ are defined
from $(t,r)$ by
\begin{equation}
R^2-c^2T^2=4r_s^2\frac{f(r)}{F(r)}, \qquad
\frac{cT}{R}=\tanh\left(\frac{ct}{2r_s}\right),
\end{equation}
where
\begin{equation}
  F(r)=\frac{r_s}{r}e^{-r/r_s}.
\end{equation}
In the Kruskal coordinate system $(T,R,\theta,\phi)$,
the metric becomes
\begin{equation}
 ds^2=-F(r)\,c^2dT^2+F(r)\,dR^2+
r^2(d\theta^2+\sin^2\theta d\phi^2),
\end{equation}
where the radial coordinate $r$ is now interpreted
as a function of $T$ and $R$.
In light of this coordinate system,
we choose a new vierbein $\dut{\tilde{e}}{a}{\mu}(x)$ as
\begin{eqnarray}
\dut{\tilde{e}}{0}{T}(x)=\frac{1}{c\sqrt{F(r)}}, &\quad&
\dut{\tilde{e}}{1}{R}(x)=\frac{1}{\sqrt{F(r)}},\nonumber \\
\dut{\tilde{e}}{2}{\theta}(x)=\frac{1}{r}, &\quad&
\dut{\tilde{e}}{3}{\phi}(x)=\frac{1}{r\sin\theta},
\label{eq:kruvb}
\end{eqnarray}
which is not singular at the horizon.
Using the original coordinate system $(t,r,\theta,\phi)$,
we find that this vierbein is related to
the static vierbein (\ref{eq:schvb})
by a local Lorentz transformation:
$\dut{\tilde{e}}{a}{\mu}(x)=\dut{\tilde{\Lambda}}{a}{b}(x)
\dut{e}{b}{\mu}(x)$, where
\begin{equation}
\dut{\tilde{\Lambda}}{a}{b}(x)=
 \left(\begin{array}{cccc}
   \frac{1}{2r_s}\sqrt{\frac{F(r)}{f(r)}}R &
   -\frac{c}{2r_s}\sqrt{\frac{F(r)}{f(r)}}T & 0 & 0 \\
   -\frac{c}{2r_s}\sqrt{\frac{F(r)}{f(r)}}T &
   \frac{1}{2r_s}\sqrt{\frac{F(r)}{f(r)}}R & 0 & 0 \\
   0 & 0 & 1 & 0 \\
   0 & 0 & 0 & 1
 \end{array}\right).
\end{equation}
Since this Lorentz transformation is a boost along the $1$-axis
parallel to the $r$-direction,
the new local inertial frame falls into the black hole when $T>0$.
To perform measurements in this local inertial frame,
the observers also must fall into the black hole.
Similarly, we choose the four-velocity of particles as
\begin{equation}
  \tilde{u}^{T}(x)=\frac{\cosh\tilde{\xi}}{\sqrt{F(r)}},\qquad
  \tilde{u}^{\phi}(x)=\pm\frac{c\,\sinh\tilde{\xi}}{r},
  \label{eq:krufv}
\end{equation}
which is not singular at the horizon.
Since $\tilde{p}^1(x)=\udt{\tilde{e}}{1}{\mu}(x)\tilde{u}^{\mu}(x)=0$,
the particles also fall into the black hole with
the local inertial frame (\ref{eq:kruvb}),
while moving in the $\pm\phi$-directions.
With the vierbein (\ref{eq:kruvb}) and
four-velocity (\ref{eq:krufv}),
we obtain the local Wigner rotation
\begin{eqnarray}
 \udt{\tilde{\vartheta}}{1}{3}(x) &=&
 -\udt{\tilde{\vartheta}}{3}{1}(x)  \nonumber \\
 &=& \pm\frac{c\cosh\tilde{\xi}\sinh\tilde{\xi}}{r}
  \left[3+\frac{r}{r_s}\right]
  \frac{\sqrt{F(r)}R}{4r_s},
\end{eqnarray}
instead of Eq.~(\ref{eq:schwig}).
Since this precession is not singular at the horizon,
the observers on the horizon
can extract the EPR correlation from the particles
without the infinite accuracy of $\Phi$.
Furthermore, they can extract the EPR correlation
beyond the horizon $r<r_s$
(until the physical singularity $r=0$).

\subsection{Bell's Inequality}
We next consider Bell's inequality using
the circularly moving particles (\ref{eq:schfv})
in the static local inertial frame (\ref{eq:schvb}).
Suppose that the spin component of one particle is measured
in the $(1,0,0)$-direction (component $\mathcal{Q}$)
or in the $(0,1,0)$-direction (component $\mathcal{R}$)
in the local inertial frame at $\phi=\Phi$,
and suppose that the spin component of the other is measured
in the $(-1,-1,0)/\sqrt{2}$-direction (component $\mathcal{S}$)
or in the $(1,-1,0)/\sqrt{2}$-direction (component $\mathcal{T}$)
in the local inertial frame at $\phi=-\Phi$.
This set of observables gives rise to the maximal violation
of Bell's inequality for the spin-singlet state
in the usual case.
However, for the circularly moving particles
in the Schwarzschild spacetime,
the degree of the violation of Bell's inequality
apparently decreases as
\begin{equation}
\langle\mathcal{QS}\rangle+\langle\mathcal{RS}\rangle+
\langle\mathcal{RT}\rangle-\langle\mathcal{QT}\rangle=
2\sqrt{2}\,\cos^2\Theta.
\end{equation}
Again, this result includes the effect of
the trivial rotations of the local inertial frames $\pm\Phi$.
To get rid of this effect,
the observers rotate the directions of the measurement
about the $2$-axis through the angles $\mp\Phi$, respectively.
That is, the spin component of one particle is measured
in the $(\cos\Phi,0,-\sin\Phi)$-direction (component $\mathcal{Q'}$)
or in the $(0,1,0)$-direction (component $\mathcal{R'}$), and
the spin component of the other is measured
in the $(-\cos\Phi,-1,-\sin\Phi)/\sqrt{2}$-direction
(component $\mathcal{S'}$)
or in the $(\cos\Phi,-1,\sin\Phi)/\sqrt{2}$-direction
(component $\mathcal{T'}$).
Nevertheless, the degree of the violation
of Bell's inequality still decreases as
\begin{equation}
\langle\mathcal{Q'S'}\rangle+\langle\mathcal{R'S'}\rangle+
\langle\mathcal{R'T'}\rangle-\langle\mathcal{Q'T'}\rangle=
2\sqrt{2}\,\cos^2\Delta,
\label{eq:relbell}
\end{equation}
due to acceleration and gravity.
Of course, local realistic theories cannot be restored,
since Eq.~(\ref{eq:relbell}) is a consequence of
local unitary operations.
This decrease means that it is a different set of directions
that maximally violates Bell's inequality.
To utilize the violation of Bell's inequality
for quantum communication,
the observers must take into account
the general-relativistic effect
arising from acceleration and gravity.
More specifically,
the spin component of one particle must be measured
in the $(\cos\Theta,0,-\sin\Theta)$-direction 
or in the $(0,1,0)$-direction
in the local inertial frame at $\phi=\Phi$, and
the spin component of the other must be measured
in the $(-\cos\Theta,-1,-\sin\Theta)/\sqrt{2}$-direction
or in the $(\cos\Theta,-1,\sin\Theta)/\sqrt{2}$-direction
in the local inertial frame at $\phi=-\Phi$.

However, in practice, it becomes difficult to
observe the violation of Bell's inequality
when an uncertainty in $\Phi$ is near the horizon.
Even if the directions of measurement are adjusted
so that Bell's inequality is maximally violated,
the error in $\Theta$ decreases
the degree of violation as
$2\sqrt{2}\cos^2\delta\Theta$.
This value must be greater than $2$
to verify the violation of Bell's inequality.
Therefore, from Eq.~(\ref{eq:err}),
$\delta\Phi$ and $r$ must satisfy at least
\begin{equation}
  \delta\Phi <\sqrt{2}\,
  \left|1+\frac{\Delta}{\Phi}\right|^{-1}.
\end{equation}
For a given uncertainty $\delta\Phi$,
there exists a radius $r_b$ ($>r_s$) such that
static observers at $r<r_b$ cannot
observe the violation of Bell's inequality
from circularly moving particles.
To see the violation of Bell's inequality at $r<r_b$,
the observers must fall into the black hole
together with the particles,
using the vierbein (\ref{eq:kruvb}) and
the four-velocity (\ref{eq:krufv}).

Using a different definition of a relativistic spin,
Czachor~\cite{Czacho97} obtained a decrease
in the degree of violation of Bell's inequality.
This decrease was caused
by the inertial motion of particles in Minkowski spacetime.
In contrast to this result,
Bell's inequality is unaffected except for
a trivial rotation in our formulation in that case.
Czachor's effect is thus different from ours.
Terno~\cite{Terno02} discussed a relation of
different choices of relativistic spin operators
to the violation of Bell's inequalities.

\section{\label{sec:summary}Summary}
We considered the EPR correlation and
the violation of Bell's inequality with accelerated particles
in a gravitational field.
Using relativistic quantum theory in curved spacetime,
we explictly derived the local Wigner rotation during
the motion of the particle.
Considering particles in a circular motion
in the Schwarzschild spacetime,
we showed that acceleration and gravity deteriorate
the EPR correlation in the directions
that are the same in non-relativistic theory,
and apparently decrease
the degree of the violation of Bell's inequality.
This finding indicates
neither a breakdown of the non-local correlation
nor a restoration of local realistic theories.
In fact, if the spins are measured
in appropriately chosen different directions,
we can obtain the perfect anti-correlation and
the maximal violation of Bell's inequality.
Our results mean that,
in order to utilize the non-local correlation
and the violation of Bell's inequality for quantum communication,
we must take account of the relativistic effect
by adjusting the directions of measurement;
otherwise, the accuracy of quantum communication is reduced.
In principle, we need information about the four-velocity and
the vierbein in order for the communication to be perfect.

Moreover, we showed that
near the event horizon even a small uncertainty in
the identification of observers' positions
results in a fatal error in identifying
the measurement direction needed
to maintain the perfect EPR correlation,
because of an extremely rapid spin precession.
In particular,
static observers on the horizon
can extract the EPR correlation from
circularly moving particles
only if they have infinite accuracy as to
their own positions.
To exploit the EPR correlation on and beyond the horizon,
the observers must choose a four-velocity and vierbein
that are not singular at the horizon,
and thus the observers must fall into the black hole
together with the particles.
This example demonstrates that
the choices of four-velocity and vierbein
are important to the ability to communicate non-locally
in a curved spacetime
using an EPR pair of spins.

\section*{Acknowledgments}
H.T. was partially supported
by JSPS Research Fellowships for Young Scientists.
This research was supported by a Grant-in-Aid
for Scientific Research (Grant No.15340129)
by the Ministry of Education, Culture, Sports,
Science and Technology of Japan,
and by the Yamada Science Foundation.


\end{document}